\documentclass[aps,prc,floats,floatfix,preprint,superscriptaddress,nofootinbib]{revtex4} 
 
\usepackage{graphicx}

\def\be{\begin{equation}} 
\def\ee{\end{equation}} 
\def\rb{\rangle} 
\def\lb{\langle} 
\def\mod{{\rm mod}} 
\begin{document}

\title{Neutron-Proton pairing revisited } 
 
\author{W.A. Friedman$^{1,2}$ and G.F. Bertsch} 
\affiliation{Department of Physics and Institute for Nuclear Theory,\\ 
             University of Washington, Seattle, WA 98195\\
$^{2}$ Department of Physics, University of Wisconsin, WI 53706} 
 
\date{February 22, 2007}

\begin{abstract} 
 
We reexamine neutron-proton pairing as a phenomenon that should be
explanable in a microscopic theory of nuclear binding energies. Empirically,
there is an increased separation energy when both neutron and proton numbers
are even or if they are both odd.  The enhancement is present at some level
in nearly all nuclei: the separation energy difference has the opposite sign
in less than 1\% of the cases in which sufficient data exist.  We discuss
the possible origin of the effect in the context of density functional
theory (DFT) and its extensions.  Neutron-proton pairing from the
Hartree-Fock-Bogoliubov theory does not seem promising to explain the
effect.  We demonstrate that much of the increased binding in the odd-odd
system might be understood as a recoupling energy.  This suggests that the
DFT should be extended by angular momentum projection to describe the
effect.

\end{abstract} 
 
\maketitle 
\label{sect:intro}

\begin{figure} 
\includegraphics [width = 11cm]{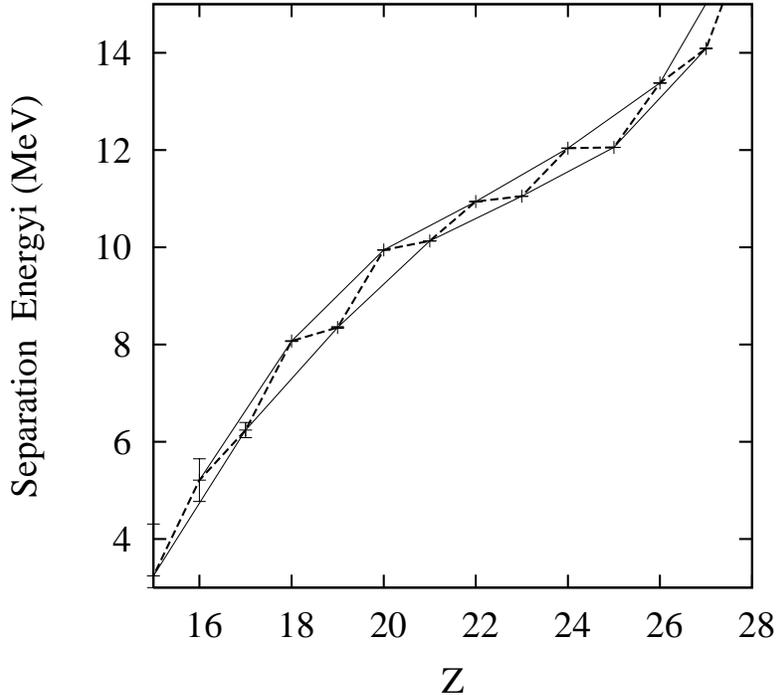} 
\caption{Neutron-proton pairing effect as seen in the neutron 
separation energy for $N=28$ as a function of proton number $Z$. 
There is a consistent offset of the separation energies of odd-$Z$ 
nuclei as compared with the average of the neighboring even-$Z$ 
nuclei. 
} 
\label{fi:1} 
\end{figure}

It has been known for a long time that the nuclear binding 
has a mild dependence on the combined even-odd parities 
of proton and neutron numbers\cite[p.171]{BM},\cite{ze75,my75}.  To see the 
effect, Fig.~\ref{fi:1} shows the neutron separation energies  for 
nuclei with neutron number $N=28$ as a function of proton 
number $Z$; the separation energy is expressed in terms 
of the binding energy as 
$S_n(N,Z) = B(N,Z)-B(N-1,Z)$.   
One sees that the neutron separation 
energies for even $Z$ are systematically larger than the 
average of the separation energies for the neighboring odd-$Z$ nuclei.   Similar behavior is found for  
proton separation energies $S_p$ in chains of isotopes. 
In that case $S_p$ is greater 
if the number of neutrons is even than when the number of neutrons is odd. 
 
To study this behavior in more detail, we examine the separation energy 
differences  
$S_{n2p},S_{p2n}$, defined as the difference between the separation  energy 
and the average for the two neighboring nuclei.  This is 
\be 
\label{S_2} 
S_{n2p}=S_n(N,Z) -\left(S_n(N,Z+1)+S_n(N,Z-1)\right) /2  
\ee 
$$S_{p2n}=S_p(N,Z) - \left((S_p(N+1,Z)+S_n(N-1,Z)\right) /2 $$ 
for neutrons and protons, respectively.  These measures were first
introduced by Jensen et al. \cite{je84}. With our notation, the 
usual measure for ordinary pairing is given by (ref.~\cite[eq. 2-92,93]{BM}) 
$$ 
2\Delta_n \equiv S_{n2n} = S_n(N,Z)- 
\left((S_n(N+1,Z)+S_n(N-1,Z)\right)/2 
$$ 
for the neutron gap, $\Delta_n$, and similarly $S_{p2p}$ gives the proton gap.  
Most earlier 
studies of neutron-proton pairing used different measures for the 
effect.  In early fits of the measured binding energies \cite{my75,mo81},  
the effect was parameterized as
$$ 
\delta \sim \, \mod(N,2)\, \mod(Z,2) / A. 
$$ 
and attributed to an enhancement in the neutron-proton interaction. 
In ref. \cite{ma88},  the parameterization was changed to one 
have an approximate $A^{-2/3}$ dependence on nuclear mass number, 
\be 
\label{delta} 
\delta =  K \, \mod(N,2)\, \mod(Z,2) / A^{2/3}. 
\ee  
In ref.~\cite{mo92} a 9-point difference formula was proposed to describe 
a neutron-proton pairing energy.  This is to be compared  
the 6-point difference formula we use in eq. (1).  We also mention the 
shell-based mass fits of Zeldes\cite{ze75}, which invoke a shell-dependent 
term similar to $\delta$. 
 
\begin{figure} 
\includegraphics [width = 11cm]{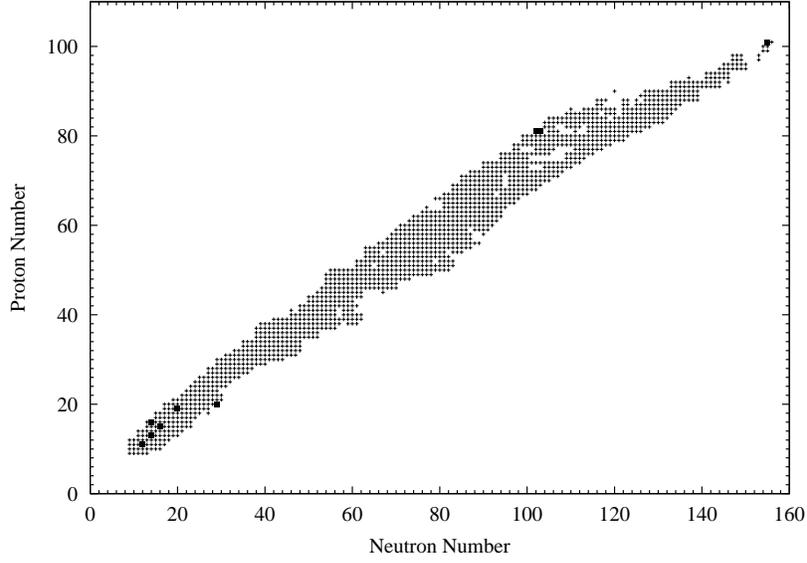} 
\caption{Nuclei with measured proton separation energy differences 
$S_{p2n}$ showing the cases (black squares) with opposite sign from  
the normal.} 
\label{fi:2} 
\end{figure}

We find that 
the signs of $S_{n2p}$ and $S_{p2n}$ are remarkably consistent across the nuclear 
mass table.  Taking the data from the 2003 Audi-Wapstra mass 
tables\cite{au03}, there 
are 1412 nuclei which have values of $S_{n2p}$ that are significant,  
i.e., 
have magnitudes larger the accumulated error in the experimental 
binding energies needed to construct the difference. 
Of these only 10 nuclei had 
a sign for $S_{n2p}$ opposite to that seen in Fig.~\ref{fi:1}.  Of the 
1448 measured proton separations $S_{p2n}$, only 9 had the opposite 
sign.  The nuclei with significant values of $S_{p2n}$ for proton separations are shown in 
Fig.~\ref{fi:2}, with the few opposite-sign cases shown as the black squares. 
The plot for neutron separations is very similar.  There is concentration
in the light nuclei near the $N=Z$ line, but no obvious pattern elsewhere.
 
\begin{figure} 
\includegraphics [width = 11cm]{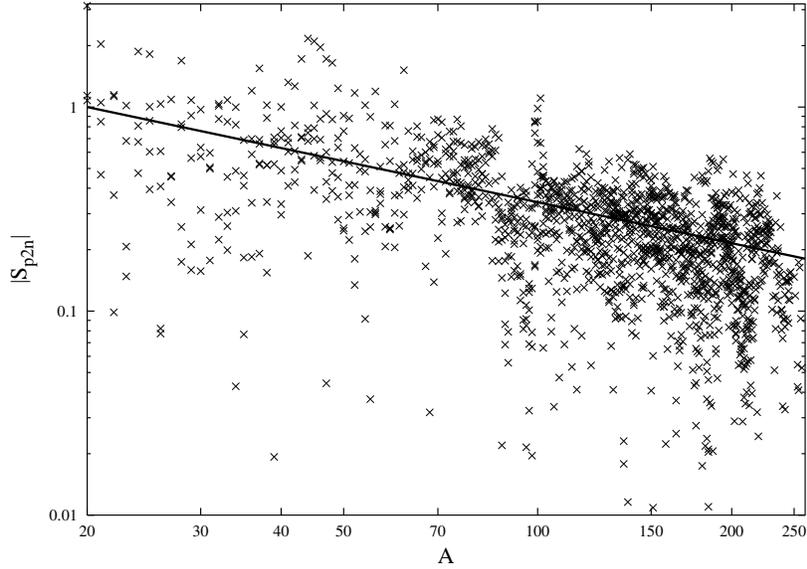} 
\caption{$\vert S_{p2n} \vert$ as a function of $A$.  The line 
 shows the $A^{-2/3}$ fit eq.~(\ref{delta}) with $K^{\prime} = 7.3 MeV$. 
} 
\label{fi:3} 
\end{figure}

 We have also examined the dependence of the magnitude of the separation
energy
differences on the mass numbers, A, of the nuclei.  There is a great deal of 
scatter as shown in Fig.~\ref{fi:3}, but the trend is consistent with an 
$A^{-2/3}$ dependence as in eq. (\ref{delta}).  
The heavy line shows a least squares fit to 
the data,  
$\vert S_{p2n} \vert =7.3/A^{2/3}$ MeV. The values for $\vert S_{n2p}\vert$ 
display a very similar distribution. 
 
\begin{figure} 
\includegraphics [width = 11cm]{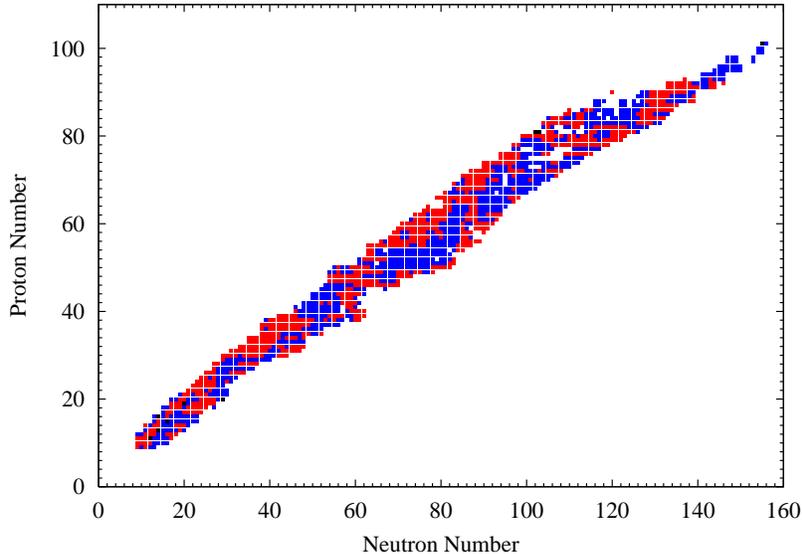} 
\caption{Chart of nuclides showing the distribution of $S_{p2n}$ values 
that are higher (red) or lower (blue) than the average trend 
$\vert S_{p2n}\vert = 7.3/A^{2/3} $ MeV. 
}   
\label{fi:4} 
\end{figure}

Finally,  we plot in Fig.~\ref{fi:4} all the measured nuclei,
distinguishing by color those whose $|S_{p2n}|$ is larger 
or smaller than the average trend.  
We  
also find a very similar pattern for $\vert S_{n2p}\vert$.
There is no visible dependence on shell closures.
 
We now turn to the question of what is responsible for the effect. 
The enhanced binding could  
arise by an increased 
attraction between an odd neutron and an odd proton.  It could also 
arise by a mechanism that produced an increased binding in a  
nucleus with even numbers for both protons and neutrons.  There is no 
way to distinguish these pictures by the observed systematics of the 
separation energy differences, because even-even and odd-odd binding 
energies contribute to the separation energy difference with the same weights. 
Both pictures are consistent with the strong similarity between the 
proton and neutron separation energy differences.  Still, it is important 
to understand the origin of the effect if one is to construct accurate 
theories of nuclear binding based on microscopic theories such as
the self-consistent mean field theory, also called density functional
theory (DFT)\cite{be03,lu03}. 
Ordinary pairing between like particles is quite well explained by the 
BCS or Hartree-Fock-Bogoliubov extension of DFT.
This suggests generalizing the HFB theory 
to allow neutron-proton pairing.  Certainly, at the $N=Z$ line neutron-proton 
pairing is on a same footing as like-particle pairing. Also, the 
neutron-proton interaction is stronger than the like-particle interaction, 
in free space.  The special effects going on near $N=Z$ are 
often discussed as the ``Wigner energy".  It is usually 
parameterized in a way that does not exhibit a neutron-proton pairing 
effect away from the $N=Z$ line and we shall consider it irrelevant to 
explain the effect.  We note again that over half of the opposite-sign cases are 
near the $N=Z$ line.  There have also been limited studies of neutron-proton 
pairing in the HFB theory~\cite{ci97,sa97}.  Typically, away from the
$N=Z$ line, condensates 
form in the like-particle sectors and prevent any pairing between 
neutrons and protons.  We therefore doubt whether the effect can be 
explained without make some extension of the usual DFT+HFB theory.
 
There is a possible mechanism that only requires a mild extension 
of the DFT.   That is to  
exploit the higher degeneracy of states in the odd-odd nucleus to 
recouple the neutron and proton more favorably.  This is easiest to 
understand in the situations where the mean-field theory 
approaches either the spherical shell model or the strongly deformed limit. 
Indeed, Zeldes and Liran\cite{ze75} may have had this mechanism in 
mind in their shell-based mass parameterization.  For the shell-model 
limit, consider even-even nucleus (N,Z) that has 
a spherical mean field.  An added neutron goes into a spherical shell 
$j_n$ with an energy $\epsilon_{j_n}$. 
Similarly, an added proton goes into a shell $j_p$.   When there are both 
added neutrons and protons, there is an addition neutron-proton  
interaction energy $\lb j_n j_p |V_{np}| j_n j_p \rb_J$ depending on 
the angular momentum of the pair $J$. The neutron separation energies 
for the nuclei with proton numbers $Z,Z+1,Z+2$ are, respectively, 
 
$$S_n(N+1,Z) = -\epsilon_{j_n} 
$$ 
$$ 
S_n(N+1,Z+1) =-\epsilon_{j_n} -<j_nj_p\vert V_{np} \vert j_n j_p>_{J_g} 
$$ 
$$ 
S_n(N+1,Z+2) =-\epsilon_{j_n} - <j_n(j_p^2)^{J=0}\vert V_{np} \vert j_n (j_p^2)^{ J=0}>_{j_n} 
$$ 
In the second equation, $J_g$ denotes the angular momentum of the 
odd-odd nucleus ground state.  The last equation gives the neutron 
separation energy for the nucleus with two additional protons.  Here the angular 
momentum coupling is determined by the three-particle wave function. 
In the spherical shell model, the two protons are coupled 
to angular momentum zero in the three-particle wave function 
$|j_n(j^2_{n})^{J=0}\rb $.  Standard angular momentum recoupling gives 
the neutron-proton interaction as  
$$<j_n(j_p^2)^{J=0}\vert V_{np} \vert j_n (j_p^2)^{ J=0}>_{j_n}= 
 \sum_{J=|j_n-j_p|}^{j_n+j_p} (2 J+1) <j_nj_p|\vert V_{np} \vert j_n j_p>_J/(2j_n+1)(2j_p+1)$$ 
 
Thus, in the shell model,  the energy of the odd neutron when the proton 
number is even is the $(2J+1)$-weighted average over the possible 
neutron-proton couplings. 
 
This value can be estimated empirically from the spectrum of the 
odd-odd nuclei as the quantity 
\be 
\label{eq:recouple} 
\delta_s = \sum_{J=|j_n-j_p|}^{j_n+j_p} (2 J+1) E_J /(2j_n+1)(2j_p+1) 
\ee 
Here $E_J$ are measured excitation energies of the levels of the multiplet 
in the odd-odd nucleus.  The quantity $\delta_s$ is thus a measure of the enhancement  
of the neutron separation energy for an odd neutron
in a nucleus with an odd number of protons. 
 
\begin{figure} 
\includegraphics [width = 11cm]{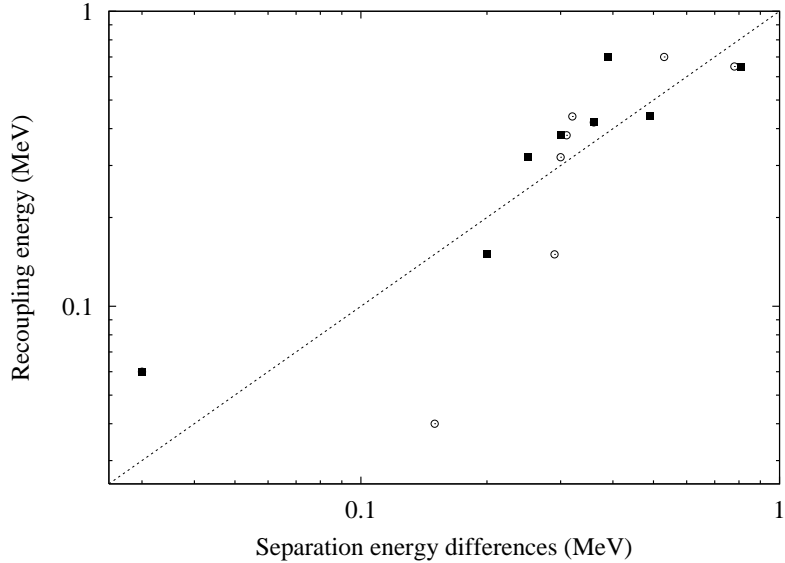} 
\caption{Scatter plot of $\delta_s$, eq. (\ref{delta}), compared with 
$S_{n2p}$ (solid squares) and $S_{p2n}$ (circles).  The nuclei plotted are 
$^{40}$K, $^{48}$Sc, $^{50}$Sc, $^{56}$Co, $^{60}$Co, $^{58}$Cu, $^{208}$Tl, 
$^{208}$Bi, and $^{210}$Bi. 
} 
\label{fi:5} 
\end{figure}

For most odd-odd nuclei, 
the recoupling spectrum is difficult to determine due to the presence
of other levels. However, near doubly magic 
nuclei it is often possible to make a spectroscopic
identification\cite{bnl}.
Some  cases where we could plausibly assign the 
members of the multiplet are shown in Table~\ref{ta:1}.   
These results are also shown in Fig.~\ref{fi:5}.  The recoupling
energy $\delta_s$ has the same order of magnitude as the separation
energy differences, and also varies from case to case in a
similar way.  However,  there is considerable scatter leaving room
for other mechanism to have a role.
\begin{table} 
\caption{ Comparison of neutron-proton pair interaction energies 
with the recoupling model, eq. \ref{eq:recouple}. Energies are in 
MeV.  The quantity $\delta_s$ is defined in eq.~(\ref{eq:recouple}).} 
\label{ta:1} 
\begin{tabular} {|cc|ccc|} 
\hline 
$N$ & $Z$ & $S_{n2p}$ & $S_{p2n}$ & $\delta_s$\\ 
\hline 
21 & 19 &0.49 & 0.32 & 0.44 \\ 
27 & 21 & 0.39 &0.53 & 0.70 \\ 
29 & 21 & 0.25 &0.30 & 0.32 \\ 
29 & 27 &0.30 &0.31 & 0.38 \\ 
29 & 29 & 0.81& 0.78 & 0.65\\ 
33 & 27 & 0.20 &0.29 & 0.15 \\ 
81 & 51 & 0.24 & 0.22 & 0.14 \\ 
125 & 83 &0.03 & 0.03 & 0.06 \\ 
127 & 81 & &0.15 & 0.04\\ 
127 & 83 &0.36 &0.36 & 0.42 \\  
\hline 
\end{tabular} 
\end{table} 

If the odd-odd recoupling is dominant to produce the effect,
it should be suppressed in deformed nuclei. 
The reason is that the coupling of the orbitals to the symmetry 
axis removes most of the degeneracy.  Only the $K$ quantum number 
remains, leaving a degeneracy between the two states $|K_p-K_n|$ 
and $K_p+K_n$ in the odd-odd nucleus.  To 
see whether the suppression is indeed present, we have examined 
the separation energies differences for strongly deformed even-even nuclei. 
We took the classification of deformed nuclei from ref.~\cite{sa06}, which used the  theoretical 
criterion that the static deformation of the nucleus be larger 
than the fluctuations about the minimum.  There are 92 nuclei
with measured $S_{n2p}$ that meet the criterion.
Fitting eq.~(\ref{delta}) to these nuclei, we find 
a slightly lower value for $K$, 5.7 MeV compared to 7.3 MeV.  Also a
larger fraction of the deformed nuclei have very small values of 
the separation energy differences:  23\% of the deformed nuclei 
have $S_{n2p}$ less than $3.7/A^{2/3}$ MeV versus 9\% for the 
other nuclei. 
The difference is not very large, suggesting that 
other mechanism beyond the recoupling effect may be needed.
For example, configuration mixing arising from the 
neutron-proton interaction might depend on whether those nucleons 
are part of a pairing condensate or not.  Such a mechanism would 
be beyond the usual DFT.
 
\section*{Acknowledgment} 
 
This work was supported in part by the Department of Energy under 
grants DE-FG02-00ER41132 and DE-FC02-07ER41457. WAF wishes to thank the Department of Physics and INT 
at the University  of Washington for their hospitality. 


\begin{thebibliography}{99} 
 
\bibitem{BM} A. Bohr and B. Mottelson, ``Nuclear Structure", (Benjamin, 
New York, 1969), Vol. I. 

\bibitem{ze75} N. Zeldes and S. Liran, Atomic Data and Nuclear Data Tables 17 
431 (1975). 

\bibitem{my75} W.D. Myers, Atomic Data and Nuclear Data Tables 17 
413 (1975). 

\bibitem{je84} A.S. Jensen, P.G. Hansen, and B. Jonson, Nucl. Phys.
A431 393 (1984).
\bibitem{mo81} P. M\"oller and J.R.~Nix, At. Data Nucl. Data Tables 26 165 
(1981).  
\bibitem{ma88} D.G.~Madland and J.R.~Nix, Nucl. Phys. A476 1 (1988). 

\bibitem{mo92} P. M\"oller and J.R.~Nix, Nucl. Phys. A536 20 (1992). 

\bibitem{au03} G. Audi, A.H.Wapstra, and C. Thibault, Nucl. Phys. 
A729 337 (2003). 

\bibitem{be03} 
  M. Bender, P.-H. Heenen, and P.-G. Reinhard, 
  Rev. Mod. Phys. \textbf{75}, 121 (2003). 

\bibitem{lu03} D. Lunney, J.M. Pearson, C. Thibault, Rev. Mod. Phys.
\textbf{75}, 1021 (2003).
\bibitem{ci97}  
Civitarese, M. Reboiro, P. Vogel, Phys. Rev. C56, 1840 
(1997).  
\bibitem{sa97} W. Satula and R. Wyss, Phys. Lett B393 1 (1997). 
\bibitem{bnl} Brookhaven Evaluated Nuclear Structure Data File,
{\tt http://www.nndc.bnl.gov/ensdf/}. 
\bibitem{sa06} B. Sabbey, M. Bender, G. F. Bertsch, P.-H. Heenen, 
arXiv:nucl-th/0611089 (2006). 
  
\end{thebibliography}
\end{document}